# RACS2: A Framework of Remote Autonomous Control System for Telescope Observation and its application

Zhi-yue Wang [a, ‡], Guang-yu Zhang [a, ‡], Jian Wang [a,b,*], Qian Zhang [a], Zhe Geng [a, ‡], Ze-yu Zhu [a], Jia-Yao Gu [b], Zhen-hao Zheng [b], Lu-cheng Zhu [b], Kun Ge [a], Hong-fei Zhang [a]

[a]State Key Laboratory of Particle Detection and Electronics, Department of Modern Physics, University of Science and Technology of China, Hefei 230026, China;
[b]Institue of Advanced Technology, University of Science and Technology of China, Hefei, 230088, China

**Abstract:** As the demand of astronomical observation rising, the telescope systems are becoming more and more complex. Thus, the observatory control software needs to be more intelligent, they have to control each instrument inside the observatory, finish the observation tasks autonomously, and report the information to users if needed. We developed a distributed autonomous observatory control framework named Remote Autonomous Control System 2nd, RACS2 to meet these requirements. The RACS2 framework uses decentralized distributed architecture, instrument control software and system service such as observation control service are implemented as different components. The communication between components is implemented based on a high-performance serialization library and a light-weighted messaging library. The interfaces towards python and **Experimental Physics and Industrial Control System (EPICS)** are implemented, so the RACS2 framework can communicate with EPICS based device control software and python-based software. Several system components including **log**, **executor**, **scheduler** and other modules are developed to help observation. Observation tasks can be programmed with python language, and the plans are scheduled by the **scheduler** component to achieve autonomous observation. A set of web service is implemented based on the **FastAPI** framework, with which user can control and manage the framework remotely. Based on the RACS2 framework, we have implemented the DATs telescope's observation system and the space object observation system. We performed remote autonomous observation and received many data with these systems.

‡ E-mail:zguangyu123@gmail.com
* E-mail: wangjian@ustc.edu.cn

## 1. INTRODUCTION

The astronomical telescopes have undergone great changes since it was first invented in the 17th century. Increasing aperture improves the light-collecting ability of telescopes. Active optics (**AO**) and other technologies improve the image quality of large-aperture telescopes. Owing to these technologies, astronomical telescopes are no longer single optical instruments, but large scientific device systems consisting of many subsystems, such as optical systems, camera systems, telescope mounts, domes, weather stations and even adaptive optics or active optics systems. On the other hand, since various environmental factors, i.e., climate, light pollution, number of clear nights and air turbulence will affect the observation, candidate sites are often selected at plateau, mountain or Antarctic for better imaging quality and higher observation efficiency. Many of these sites are far from cities and not suitable for human survival, a manual control simple controlled system can hardly survive such environment. The system has to control every subsystem, perform the observation automatically, and be operated remotely.

A. J. Castro-Tirado's paper[1] introduced the concept of *Robotic autonomous observatories*. In this paper, the telescope systems are categorized into four levels by the degree of automation. 1. The Automated Scheduled Telescopes, 2. The Remotely Operated Telescopes, 3. Autonomous Robotic Observatories (ARO), and 4. Robotic Autonomous Observatories Networks (RAO). According to this classification, many observatories are between ARO and RAO. They can achieve some level of robotic observation, such as selecting the next observation target according to plans, changing the observation state according to sensor information and weather, or sending notification to operators while automatically changing the observation state. As the conclusion of this paper, the future is the era of robotic astronomical network.

In order to achieve such goal, the control software of the observatory is critical. As shown in Figure 1, the control software of a typical modern autonomous robotic observatory consists of 4 subsystems[2], the Telescope Control System (TCS), the Instrumentation Control System (ICS), the Observation Control System (OCS), the Data Handling System (DHS), and the User Interface(UI). The OCS is the center of control system, it manages the resources and schedules of the whole telescope. The OCS is the top layer of the control system and other subsystems. The TCS controls the mount to point, track, and calibrate, it also controls the telescope focuser system. For more sophisticated optical telescopes which are equipped with Adaptive Optics System or Active Optics System, TCS also needs to control these systems. The ICS controls the focal plane instruments, including cameras, spectrometers, filters, shutters, focusers, and diffusers, etc. The DHS stores the data and provides corresponding access interfaces.

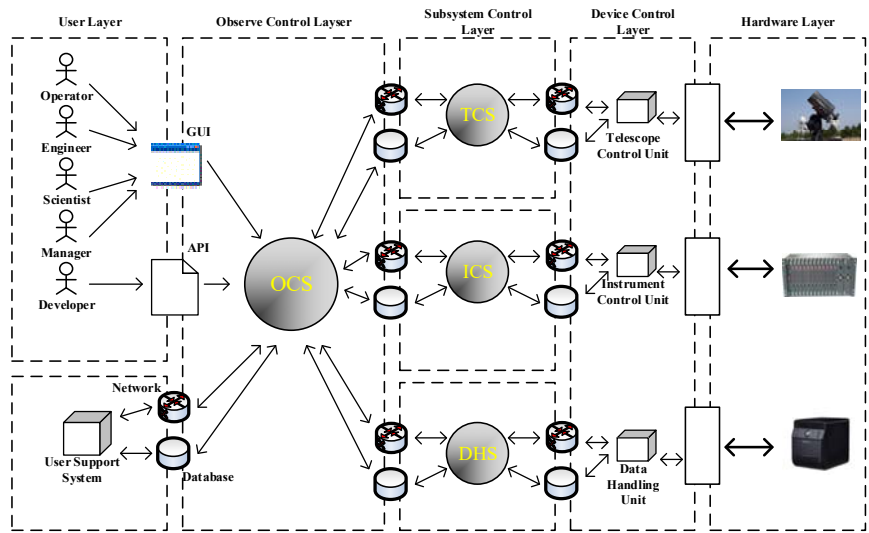

Figure 1 Telescope observational and control system framework

For large telescopes, the sophisticated requirements of system control make a single computer unable to meet the requirements. Therefore, distributed architectures are needed for large automatic telescopes. Such architecture consists of serval components, each of them is a separate program running on each node. Message brokers are used to communicate between components. The message brokers are a kind of software that delivers messages in a distributed system. The message brokers provide easy-using interfaces, so as to keep users from all the complexity and details of underlying communication mechanism. Many message brokers support several communication modes, such as point to point mode, pub/sub mode etc. Moreover, some message brokers provide more sophisticated function such as load balancing, message routing, Quality of Service (QoS) control, etc.

Typical message brokers include these types:
- Based on Remote Procedure Call (RPC): gRPC, JavaRMI
- Based on Object Request Broker (ORB): CORBA[5]
- Based on Message Queue: ActiveMQ, RabbitMQ
- Based on Data Distribution Service (DDS): OpenDDS

For telescope observation and control systems, most widely used message brokers include CORBA, DDS, Message Queue. Also, there are some customized message brokers that can't be categorized into any of these kinds (Such as the RTS2 framework who uses a socket-based communication mechanism with a set of customized protocol). Typical techniques are shown in Table.1

Table.1 Comparison of several common techniques

| Architecture | Pros | Cons | Product | Application |
|---|---|---|---|---|
| ORB | Independent of Language and OS<br>Rich features | Complex<br>Many impractical features | TAO | ALMA[6]<br>LAMOST[8] |
| DDS | Data integrity assurance<br>Rich features<br>Standardization | Complex | openDDS<br>openSlice DDS | LSST[11] [12] |
| Specific Custom Protocol | Flexibility | Compatibility<br>Poor device support | RTS2<br>CSW | BSST[13]<br>TMT[15] |
| Message Queue | High performance<br>Flexibility | Poor standardization<br>Does not have a standard structure<br>Lake of QoS support | RACS2 | DATs<br>GMT[14] |

Most big telescopes use general message buses. For example, Keck telescope uses the distributed control framework based on the EPICS which integrate various components running on different platforms[10]. Another example is the LSST, which uses OpenSlice DDS framework and its own customized message protocol to construct an observation system[11]. The TMT developed the *TMT common software* framework[15], which can be categorized as a customized protocol. The GMT control software uses Domain Specific Language, but as mentioned in this paper[14], the implementation is based on general message queues. The author's laboratory has used some of the technologies mentioned in the table above on big observatories, i.e., we implemented the LAMOST observation control system using CORBA. All these control systems are deeply customized for their own specific usages, it is almost impossible to transplant the control system of one to another.

Among small telescopes, the RTS2[19] is popular. RTS2 is developed for Linux platform. It adopts a centralized architecture: the *Centrald* service is the key component and provides functions such as naming service, node status monitoring and so on. RTS2 uses a set of its own communication protocol (RTS2 protocol) to interchange command between components and execute the corresponding task accordingly. Since CORBA and DDS are too complex for small and medium-sized telescopes, we use the lightweight RTS2 in the small aperture telescope in Antarctica[13]. Although RTS2 is widely used, it has some drawbacks and limitations. Extending RTS2 is very hard, users have to climb a very steep learning curve to understand the RTS2 class inherit mechanism, and all the extension code must be written in C/C++, other programming languages are not supported. Furthermore, RTS2 must maintain the *CentralD* service, if the *CentralD* service failed, so will the whole system. This can be a large problem for complex telescope systems, since distributed systems may encounter partial failure. When such failure occurs (such as the *CentralD* crashed and other service remain intact), we hope the rest of the system can remain useable with some basic function, and RTS2 can't achieve such goal.

To solve these drawbacks, the Remote Autonomous Control System $2^{nd}$, RACS2 is designed and developed. The RACS2 is scalable for different telescope systems, from the small amateur telescope systems to the large scientific telescope system consisting of subsystems. Typical communication delay (Including serialization and deserialization) between RACS2 components is down to tens of milliseconds. The observation task and instrument control utilities are fully decoupled and the observation tasks can be dynamically scheduled. The RACS2 can be deployed on most Linux and Unix distributions, and the binary release is available as a 10MB zip file. The RACS2 also provides binding of multiple programming languages (including Python) and other frameworks. The interfaces of the RACS2 are consistent and easy understanding. These features speed up the development of control system and help the user to further extend the control system.

## 2. SYSTEM ARCHITECTURE

### 2.1 LAYER OF OBSERVATION NETWORK AND RACS2

Modern astronomical observation systems, such as Stellar Observations Network Group, *SONG* [16], and the SiTian project[17], have shifted from single telescope to telescope network. The typical software system architecture of such an observation network is shown in Figure 2. The control framework is divided into two parts, the control system of a single observatory (Abbreviate as observation node, As shown in the right part of figure), and the managing platform (As shown in the left part of figure). Each observatory node can be further decomposed into three layers:

Layer.1 Device / data source control layer. This layer interacts with hardware directly and encapsulates the detail of low-level operation.

Layer.2 Observation control and service coordination layer. This layer will interact with Layer.1, and it will implement functions such as observation task execution, data collecting, log analysis and system diagnosis, data processing, etc.

Layer.3 Interface layer. This layer provides Graphic User Interface (GUI) and the Application Programming Interface (API). Managing platform will interact with observatory nodes via the interfaces provided by this layer. Human users can also use the GUI provided by this layer.

The managing platform gathers information from and send observation task to each observatory node of the observation network. Managing platform will analyze the data and store these data into persistence storage. The managing platform provides GUI for user to interact with it.

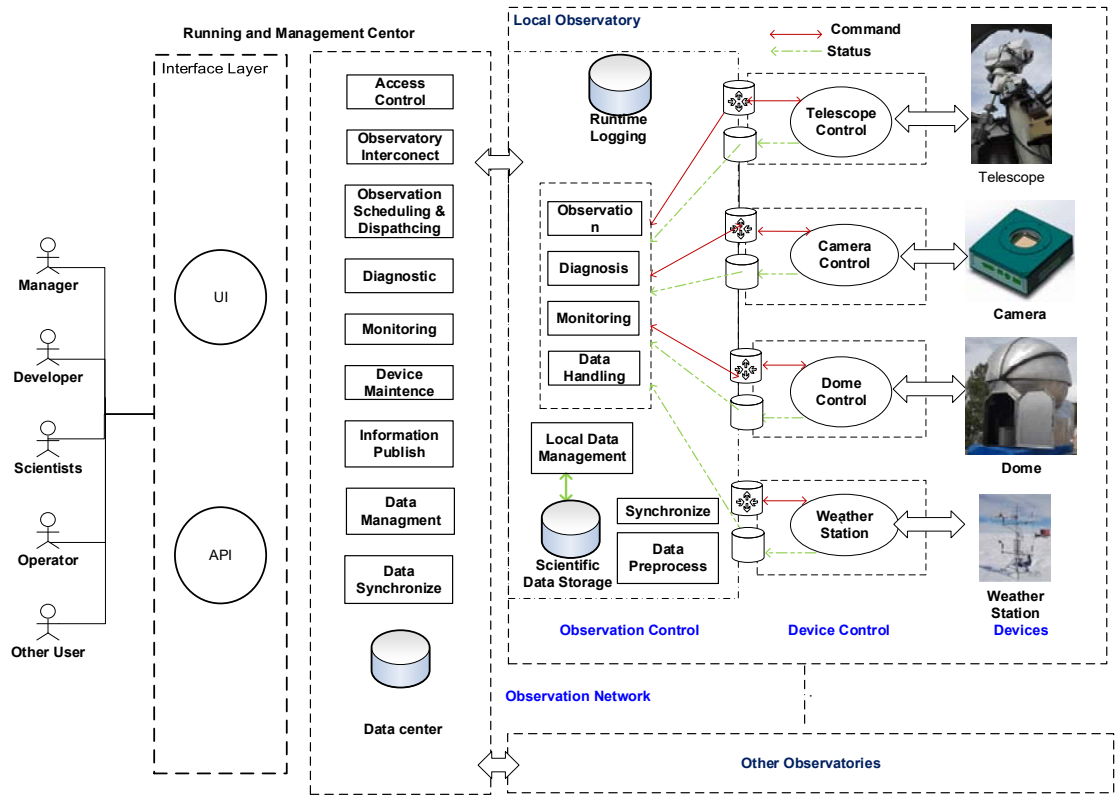

Figure 2 System Architecture

The RACS2 focus on the distributed control in a single telescope observatory. It draws the experience from frameworks such as RTS2 and EPICS, and assimilates the idea of de-centralization and components auto-discovery. A lightweight message bus is used by RACS2, and the framework specifies the remote procedure call (RPC) interfaces between components. The entire framework uses a layered architecture, as shown in Figure 3. The system consists of two layers: **RACS2 Common Lib** and **RACS2 Components**.

**RACS2 Common Lib** is a C++ shared library that implements the basic functions and interfaces of RACS2. All **RACS2 components** are developed based on the **RACS2 Common Lib**. **RACS2 components** consist of device components and service components. Each component is an independent executable program, together they make up a network of components. This scheme allows components to run on different devices in a local area network, such as embedded industrial computers of large telescopes, which make the development of control system easier.

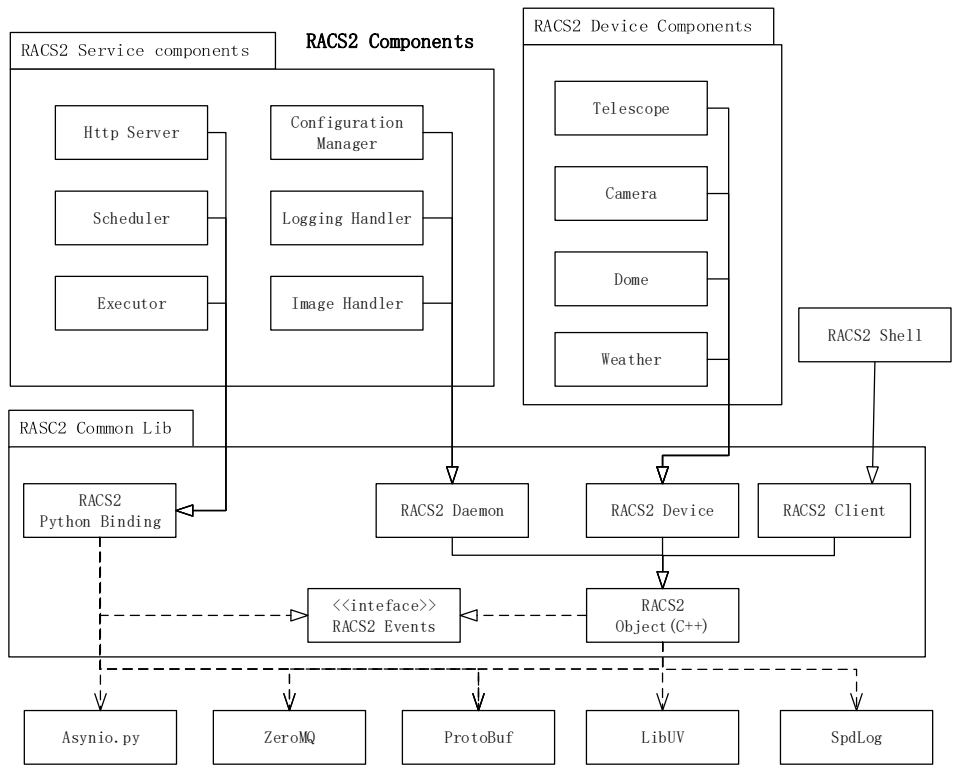

Figure 3 Conceptual Classes of RACS2 Framework

**2.2 RACS2 MESSAGE MECHANISM**

The Message Bus of RACS2 is shown in Fig.4.

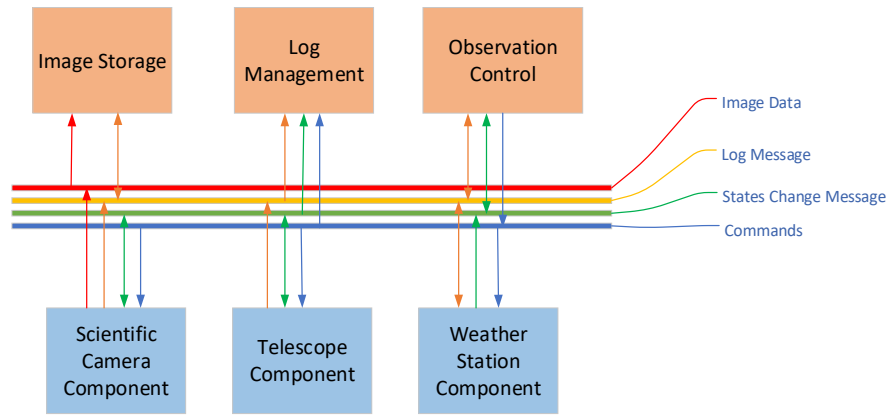

Figure 4 RACS2 Message Bus

The message bus of RACS is implemented based on ZeroMQ[22] and Google Protocol Buffer (ProtoBuf) [3]. The RACS2 implement event-driven mechanism based on LIBUV[4], inside the event-loop it uses the encapsulated socket interface provided by ZeroMQ to communicate between components. Both peer-to-peer mode and broadcast mode are implemented. Message serialization and deserialization are done by ProtoBuf. RACS2 provides fine-grained grouping strategies to manage these components, by default all the components will enter a main group, inside which the state change message of each component will be broadcast, and user can organize components with custom groups.

Local Area Network component discovery mechanism is implemented with beacon mode. Figure 5 shows an example: Component2 starts at T0, and it broadcasts its IP address and ZMQ Port. An already started component, Component1, captures this message and tries to establish connection with Component2. At T1, connection between Component1 and Component2 is successfully established, and these components send *HELLO* messages to each other. On receiving *HELLO* message, the component replies a *INFO* message, inside the *INFO* message is the meta information consists of the component's name, type, network address, properties etc. When receiving the INFO messages, the component will append the information to its *peer_list* (further described in section 2.3). By the time T2, this enter procedure finished, and both components are aware of each other. Same procedure can happen between the newcomer component and other components if there exists any.

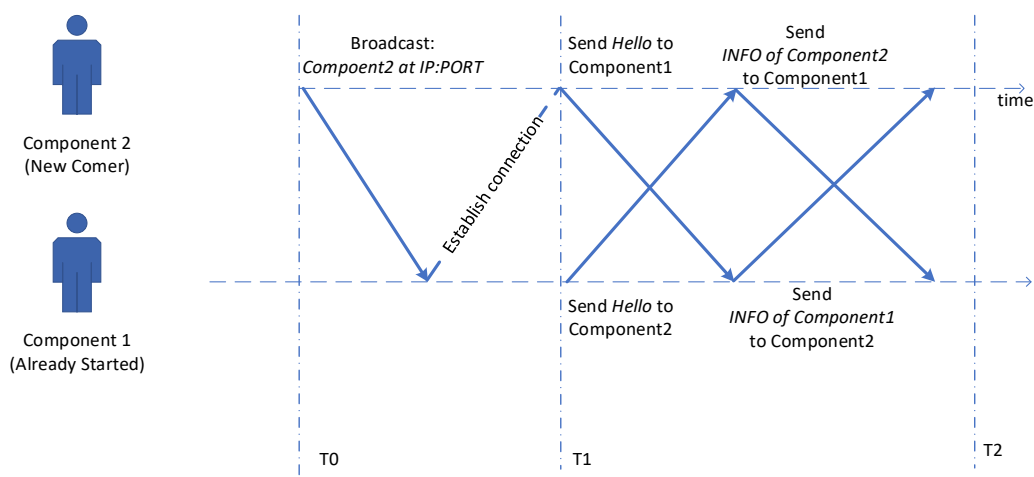

Figure 5 Demo Diagram of component discovery mechanism. Communication between Component1 and Component2 right after Compotnent2 started at T0

This mechanism improves the scalability of the system and makes it possible to expand the system online without modifying the code and configuration files. Compared with the architecture of polling each node by the central node, the decentralized architecture saves network traffic and resources.

In the observational system, each instrument or observation service is abstracted into a RACS2 component, and each *Property* of the component maps to the corresponding attribute of instrument. Operations towards *Property* trigger the operations towards hardware or function of the service. Such operations are done by sending messages to the component. Each message contains a type information in the header, there are three types of messages:

*ChangeProperty* messages and *PropertyChanged* messages identify the operations towards *Property* of components, it includes the name of the *Property* and its target value. When receiving a *ChangeProperty* messages, the component will normally trigger some corresponding procedure, such as operation towards instrument or a software function, and broadcast a *PropertyChanged* message to notify other component that the procedure is done. Figure 4 shows an example, the target RA and DEC of a telescope mount are abstracted into RA Property and DEC Property respectively, so when users want to point some certain sky area, they can send a *ChangeProtperty* message to the telescope mount component, and the component will do the underlying job such as sending command via serial port. When the rotation is completed, the telescope mount component will broadcast a *PropertyChanged* message to every component in the group.

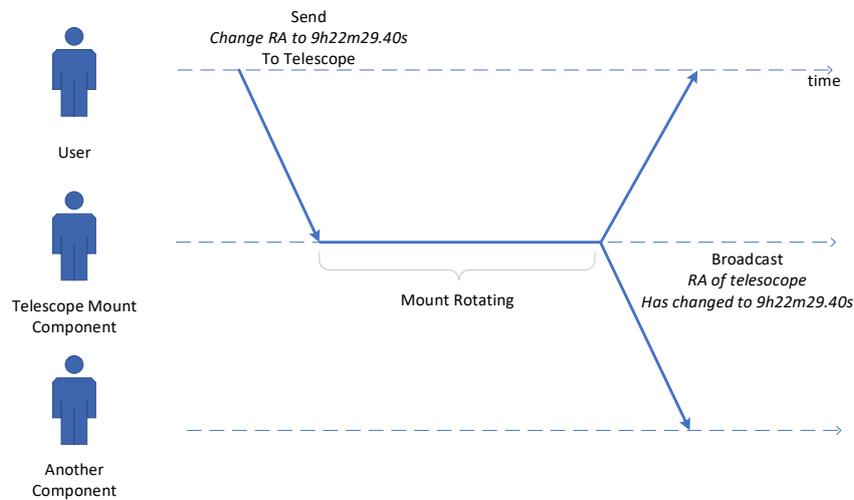

Figure 4 An example of Property Change message

*Log* messages contain some log information, the log service captures these messages and records them.

*Command* messages can be used to trigger some complicated procedure. A *command* message contains a command name and some arguments. For example, the initialization procedure of a scientific camera contains many arguments, such readout speed, gain level, cooling configuration and so on, user can define a *command* message to handle it. Sometimes, same work can be done with a group of *Property* messages, but that introduce complexity and errors, and is less intuitive than a *Command* message.

The messages typing system facilitates the development of system service components, such as logger, executor and scheduler.

**2.3 RACS2 CLASS AND EVENTS**

RACS2 is designed following the object-oriented paradigm. Users can directly inherit Daemon and Client classes to implement their own component. For some even more special requirements, users can as well as inherit Network, Logger, and other tool classes. This design enhances the scalability of the system. Figure 6 is the class diagram of RACS2 framework. As the figure shows, **EventCallback**, **Daemon** and **Client** are the base classes used by other classes, the derived classes are the tool classes mainly used to realize specific functions.

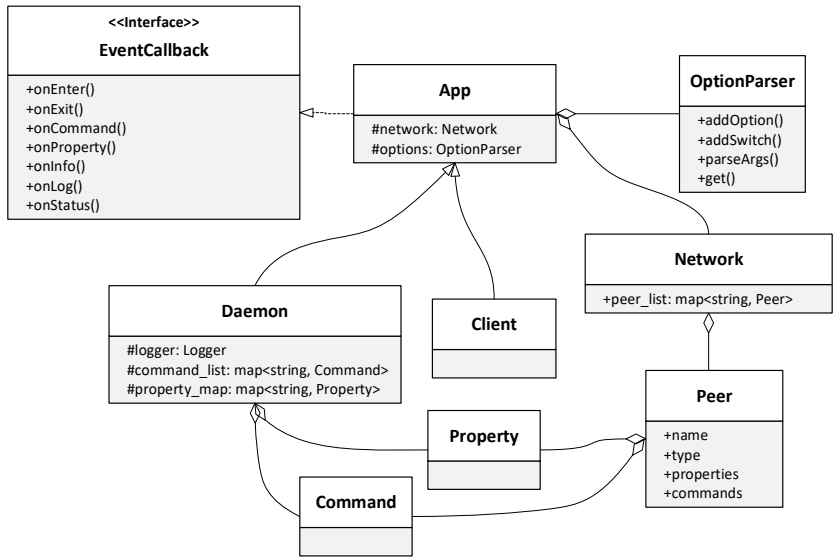

Figure 6 Class diagram of RACS2 framework

Event-driven is a key feature of RACS2, and the event interfaces are defined by the **EventCallback** class. Each interface of **EventCallback** is corresponding to an event type. The derived classed do not need to implement all event interfaces. Since these interfaces are merely a CPP function, user can call them manually and check the result, so the unit test for each derived class can be implemented easily (compare with some complicated RPC call).

We use libUV to build an event loop, with this event loop, the callback functions mentioned above will be triggered when corresponding event happened. *onEnter*, *onJoin*, and *onExit* handle the component entering and leaving procedure as we described in section 2.2. Message handling is likewise, when received a message, component will read the header of message to indicate its type, and call *onCommand*, *onLog* or *onProperty* to process *Command*, *Log*, and *ChangeProperty* messages respectively.

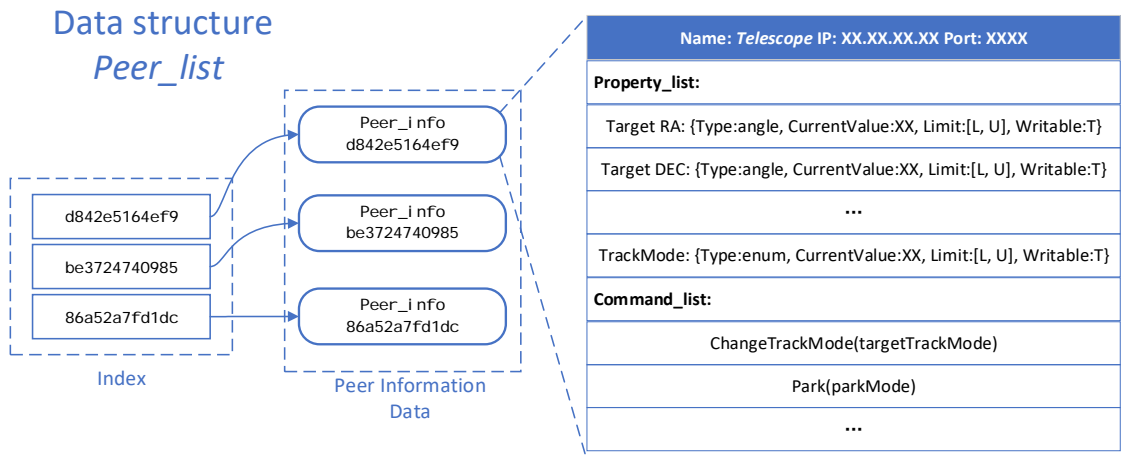

Figure 5 The peer_list data structure

Each component holds a *peer_list* to record every components in the network. The structure of *peer_list* is shown in Figure 5. As shown in the figure, the *peer_list* is an indexed data structure that records the information of every known component. The index of *peer_list* is the UUID of corresponding component. Recorded information includes the value and meta data of each *Property*, the available *Command*, and other meta data (like IP address and Port). When the component received a *PropertyChanged* message, it will update the value recorded in *peer_list*. As described in section 2.2, the peer_list is updated every time components enter or leave the network. By storing the value of *properties* inside *peer_list*, the component can access these values without querying them from network, and therefore saved some network traffic.

## 2.4 CONFIGURATION MANAGEMENT

Configuration management of RACS2 is based on configuration files. RACS2 configuration management system has the function of searching, loading, and automatic updating the configuration files. Each component maps to a configuration file named after the component name. In addition, there are independent configuration files such as EPICS component mapping file, database configuration file, FITS header mapping file and site information.

RACS2 provides a tool class called **ConfManager** for configuring operations. The structure of configuration of RACS2 is show in Figure 6, configuration data are grouped into several profiles, and inside each profile are many key-value pairs. The **ConfManager** will read the configuration files store in the file system (by default, the path is /etc/racs2 for linux system), and load all the key-value pairs recorded in these files. After the initialization, other components can access each configuration item by its profile name and key name from The **ConfManager**. The **ConfManager** also allows other components to update certain configuration item and store the new value into the configuration file.

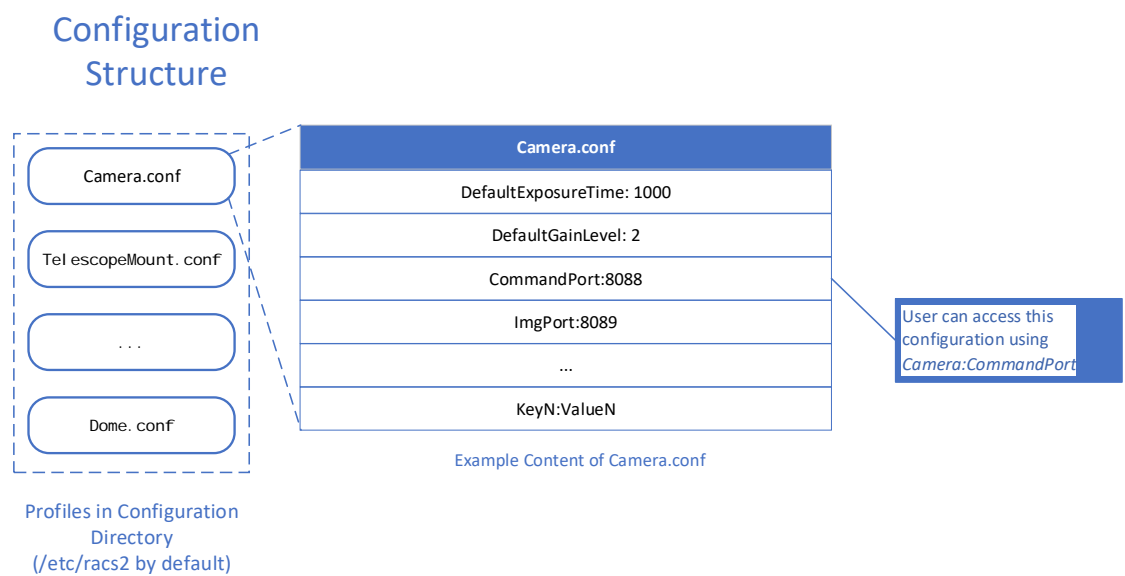

Figure 6 Structure of configuration files

A typical usage of the **ConfManager** is to store the choice of FITS[21] headers. FITS headers are often used to record many important meta data, such as exposure time and gain level, into the image data files. For some custom observation tasks, scientists may want to use some custom FITS headers. We use the **ConfManager** to implement such request: all the needed FITS headers and their attributes are stored in the FitsHeader.conf. The **ConfManager** will load this file and read the configuration items from it. When the camera component has finished an exposure task and is creating the FITS file, it will ask the

**ConfManager** about which custom headers are needed and where to find their values (normally from other components), and then write these FITS headers according to the information provided by the **ConfManager**.

A constrain is that each observation system should start one and only one **ConfManager**. This is because we want to avoid the race condition between different **ConfManager**s. On the other hand, some components may rely on the **ConfManager** to get some configuration items so one single **ConfManager** is necessary. Anyway, a hot restart or migration from one device to another of **ConfManager** is acceptable.

**2.5 INTERFACES TO OTHER SOFTWARE**

We developed a set of python binding names **Pyles**, users can implement RACS2 component in python with **Pyles**. **Pyles** is implemented based on C++ pybind11 library. Binded interfaces include the following C + + types: *EventCallback, Peer, Property, Network*, and ProtoBuf types. In the implementation of *pyles*, event loop is implemented with Python's own **asyncio** library. **Asyncio** provides asynchronous I/O operation, system signal control, idle work and other functions in Python components; In addition, **asyncio** also provides support for Python coroutines. The log class is implemented based on Python's logging module. It is compatible with the C + + version of the log class and will also automatically send logs to the message bus.

Python is more suitable for rapid development. Therefore, many RACS2 components are implemented based on Python language with *pyles*, such as log service, executor and diagnostic system.

Another important interface is the EPICS-bridge. Since we have developed many control softwares based on EPICS, we want to make use of these softwares. EPICS-bridge is a protocol parser, it listens to the zeromq message queue of RACS2, and interprets the message to EPICS Channel-Access, and also interprets the CA message to zeromq message queue of RACS2. With EPICS-bridge, EPICS-IOC can be accessed from the RACS2 components.

**3. RACS2 SYSTEM COMPONENTS**

In order to make RACS2 available for more users, we have done a lot of work on interface adapting, user GUI and automatic scheduling.

**3.1 OBSERVATIONAL CONTROL AND SERVICE COORDINATION**

The automatic observation is a key feature of automatic observatory control system. The automatic observation process is usually divided into two parts, one is the plan dispatcher, the other is the command executor.

In many observation frameworks, the design of automatic observation is often not flexible enough to meet the requirements of complex telescope applications. For example, RTS2 has a set of finite state machine to implement automatic observation, target objects are organized into a xml file, and the RTS2 executor will follow this xml file to perform the observation. Anyway, it's difficult for human user to write and read XML files, and RTS2 cannot support observation other than sequential observation, if you want the framework to automatic find next target to observe, you will have to implement such feature manually. We want to solve these drawbacks by introducing a set of more powerful observation executing mechanism into RACS2.

Instead of using XML files to record observation plans, RACS2 uses python script to describe observation plans. Inside the observation scripts, user can access every component in the network with RACS2 python interface. Python language intrinsic features are support in observation scripts. For example, *async* and *await* is usable, so user can perform asynchronous waiting in the script. For the safety consideration, we will check the *import* functions. Compared to the XML file method, python scripts can describe much more complicated procedure, including judging, branching. User can even use the python calculating libraries and databases interfaces inside the scripts.

For simple observation, such as repeatedly pointing a certain sky area, start tracking and take picture, RACS2 also provides some helping utilities. User can specify arguments for the script. And upload a CSV file with a list of arguments. RACS2 will automatically execute the script with each set of arguments in CSV file.

```python
# Acquire the parameters
ra = float(args[0])
dec = float(args[1])
exptime = float(args[2])

# Acquire the components
cam = racs2.get_peer("CAM1")
tel = racs2.get_peer("TEL1")

# Set the properties of each components
racs2.setprop(cam, "EXPTIME", exptime)
racs2.setprop(tel, "TargetRA", ra)
racs2.setprop(tel, "TargetDEC", dec)

# Tell the telescope to move to the position
racs2.sendcmd(tel, "move")

# Wait until the telescope point to the position and start
tracking, with a timeout of 30second
await racs2.wait_status(tel, "IDLE", 30)

# Send exposure command, at the same time the telescope is tracking
the object
racs2.sendcmd(cam, "expose")
# Wait until the exposure finish
await racs2.wait_status(cam, "IDLE", exptime+30)
```

List above is a typical observation script we used in DATs project. This script shows features of RACS2 observation scripts. At the beginning of the list is arguments capturing. *ra*, *dec* and exptime are the arguments passed to the script. The *racs2* python library is injected into the context at the beginning. With *racs2.setprop*, user can set the property of another component, and with

*racs2.sendcmd* user can send command to another component. The *racs2.wait_status* function allows the users to asynchronously waiting for a condition to satisfy before continue execution.

The scripts are store inside an observation database. The schema of the observation database is shown in Figure 7. Each observation plan is recorded in *plan* table. Every plan has a *script_id* which points to the observation *script* it uses, as well as the parameters which store in the *plan_args* table. The executor will check the validity of parameters at run time according to the information in *script_args* table.

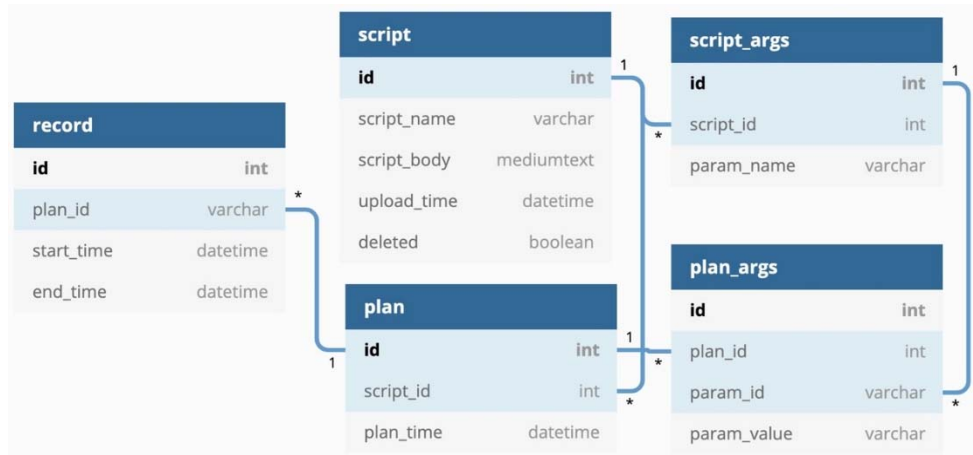

Figure 7 Database schema of observation database of RACS2

RACS2 uses two system components to conduct the observation, the **Scheduler** component and the **Executor** component.

The **Scheduler** component fetches the observation plan from database and sends the task to the **Executor**. The **Executor** will execute the observation script. When the observation plan is complete, the **Executor** can notify the **Scheduler** to fetch next observation plan.

**Scheduler**, the task management component, builds a set of state machine to maintain and schedule the execution status of each observation plan. The state transition diagram is shown in Figure 8. There are two execution modes for the observation task: step execution and periodical execution. For step execution mode, the scheduler will wait for next plan after each plan finished, so that it will perform several plans according to database. For repeat mode, the scheduler will repeat the plan until timeout. The **Scheduler** can be paused or resumed manually. The scheduler will always check the observation database for the next plan and send the script together with parameters to the **Executor**.

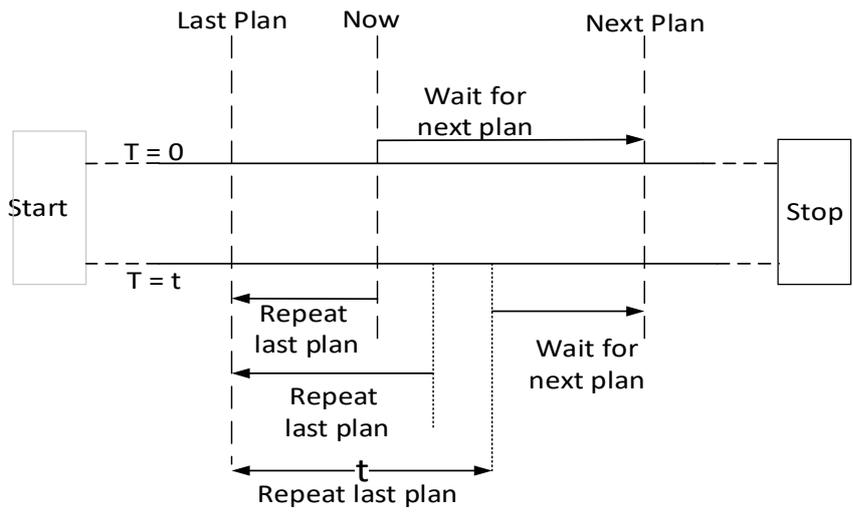

Figure 8 State transition of two work modes of the Scheduler

The **Executor**, on the other hand, will executes the scripts sent by the **Scheduler**. The **Executor** maintains state machines to describe the execution state of task. There are two execution states: IDLE and RUNNING as shown in Figure 9 Task execution state transition of the Executor.

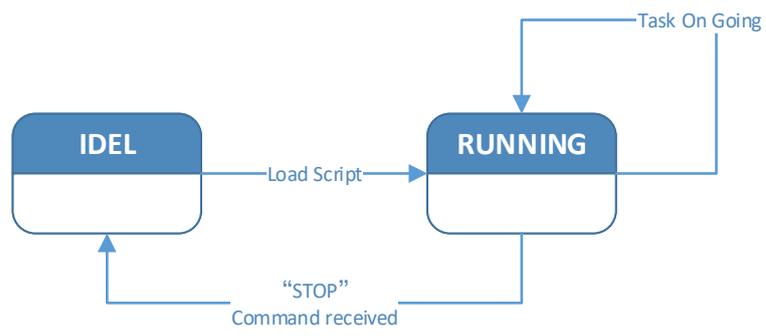

Figure 9 Task execution state transition of the Executor

The **Executor** makes use of the asynchronous characteristics of Python, the execution of task is implemented based on coroutine of *Ascynio* library. When the **Executor** receives a script, it will check the grammar and delete unsafe operations by checking the *Abstract Syntax Tree* of the script, and create a coroutine to execute the script. Since the coroutine will not block the **Executor** thread, the **Executor** can still receive other command, such as *stop* command.

### 3.2 LOG MANAGEMENT AND ITS VISUALIZATION

RACS2's log system can be divided into three parts: log collection, log storage and log visualization. The framework of log system is shown in Figure 10.

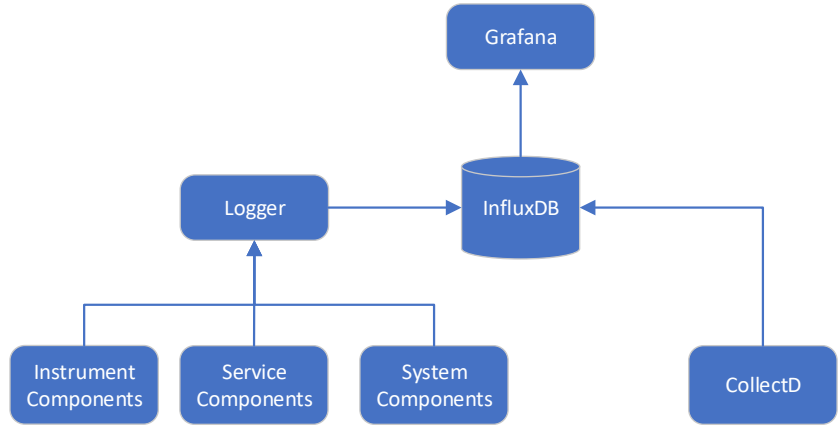

Figure 10 Framework of Log System

RACS2 provides the **Logger** component to capture state switch of components, log requests from other component and the hardware and system running information. The logger will organize this information and save into InfluxDB[24]. Different from general relation databases, the **InfluxDB**, as a Time Series Database (TSDB), contains a timestamp as the primary key of all tables. Therefore, it is an ideal storage backend for logs, and it helps to analysis, display and diagnosis the data.

Log visualization is implemented based on the **Grafana**. The **Grafana** can obtain data from database and visualize it on the dashboard in real time. RACS2 also uses the **Grafana** to send user notification when a variable exceeds alarm threshold.

### 3.3 FAULT DIAGNOSIS AND ALARM COMPONENT

Two fault diagnosis mechanisms are provided in the RACS2 framework: one is fault monitoring within a single component, the other is the alarm component which monitors the software and hardware exceptions.

The internal failure monitoring mechanism within a single component comes from the design of the component. Each RACS2 component has an error state in addition to its normal state. Error states include three states: NOERROR, NOTREADY, OUTOFRANGE. In addition, each component can define a unique variety of error states. For each subdivision state of the component, each integer type and floating point classType Property contains two attributes: *min_limit* and *max_limit*, which describe both thresholds of a property; once a variable exceeds its range, an OUTOFRANGE event is triggered, and the error information is written to the log.

The alarm component works as a service. The alarm component will acquire information from other components and find out if there are some exceptions. The rules of finding exceptions are defined by users, and the alarm component uses a python-based expert system to process these rules. If an unrecoverable failure occurs, the alarm component stops telescope observation and automatically sends notifications to users. Common recoverable failures are usually due to a mistaken command or weather factors, the alarm component will stop the observation at these times, and do some user defined recovery process.

## 3.4 GUI INTERFACE

The Graphic User Interface (UI) of RACS2 is implemented with B/S architecture. RACS2 choose to use the web based GUI for two reasons: The first reason is that Web based UI can be accessed from any device (such as ceil phone or laptop), so the users do not have to deploy the UI by themselves. The second reason is that the Web-Base UI make remote control more simple, the backend (Server) need to be deployed alongside the instruments, while the frontend(GUI) can be access remotely, so the control can be done far away from the observatory.

For RACS2, the backend is a http server based on FastAPI. The server APIs follow **RESTful[25]** discipline. The frontend is based on the Vue framework. The GUI consists of four sub-pages, integrates functions of device monitoring, device control, observation tasks management and user management. Figure 11 and Figure 12 show the task management page and the user management page. User can editing the observation scripts, store the scripts into database, and trigger an observation task manually with the frontend.

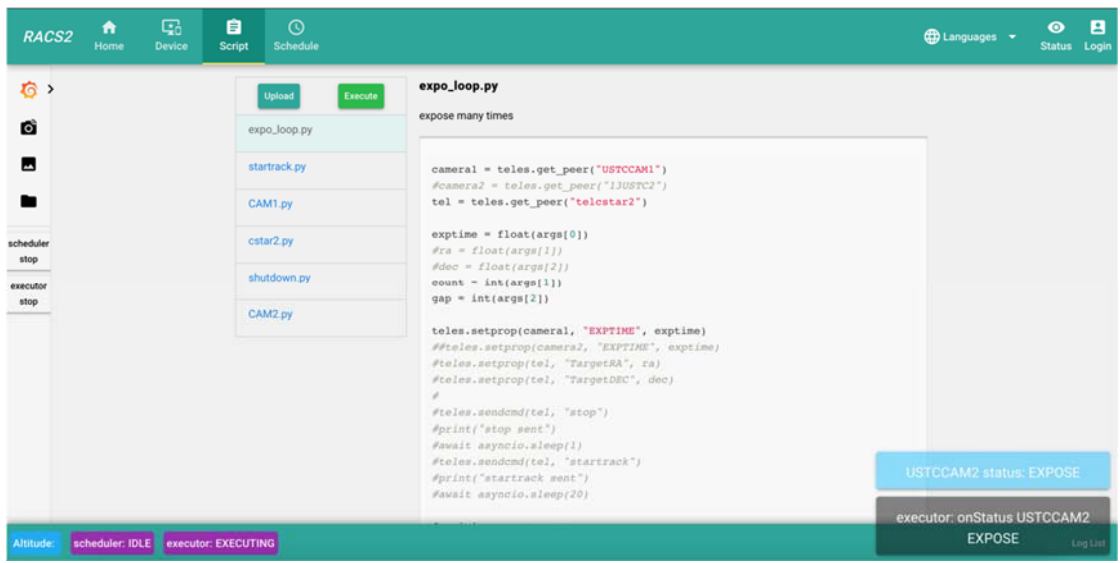

Figure 11 Script management page

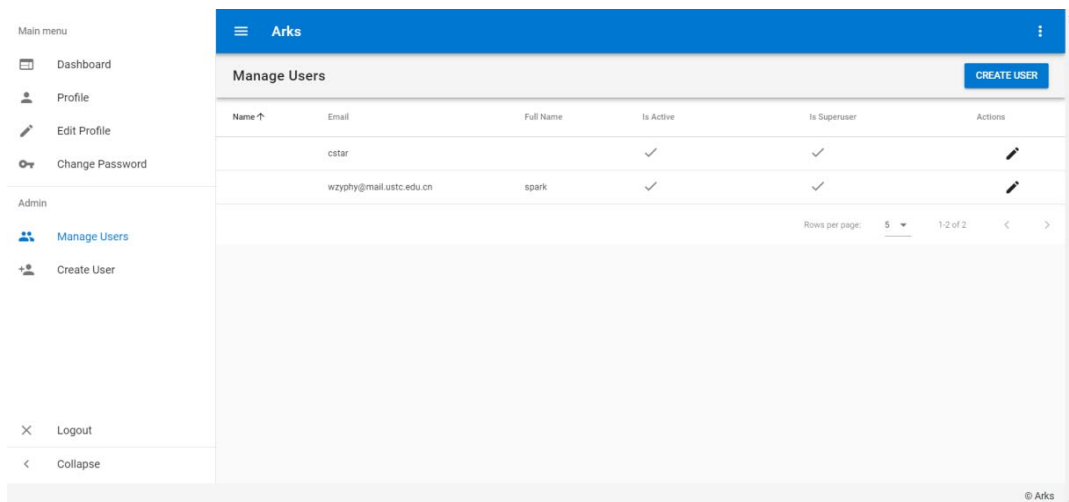

Figure 12 User management page

For security reasons, the backend has a permission hierarchy system. Users must authenticate their identity with a username and a password before accessing the GUI. Important operations such as editing the observation scripts and reading the database also need authentication. New users can be added via the user management page.

The web interface meets the requirements of observation and telescope diagnostic. Users can complete the observation task only by using the web interface remotely.

## 4. TEST AND ITS PERFORMANCE

### 4.1 RACS2 TEST

Tests and benchmarking are critical for program quality. Unit test is used to test each function in a program. The RACS2 project integrates the Google Test framework[27] to implement the unit test for functions and classes. Python API and all Python based components are tested with pytest framework. We also conduct performance test to measure the performance of RACS2 component network.

The entire project is managed with TeamCity, a Continuous Integrated (CI) platform. When new code is submitted, the server automatically compiles, publishes, and tests the project, and generates corresponding reports and software package.

### 4.2 PERFORMANCE TEST

In order to measure the performance of RACS2 framework, we measured the relationship between messaging latency and concurrency/message size of RACS2 under Local Area Network (LAN) and Link Local (LO) environment. Two components, *Server* and *Client* are started to perform the test. The *Client* component will send messages of exponential series size to the Server to measure the latency.

We also let the Client to send multiple messaging requests continuously to simulate the concurrent message request, this method is shown in following figure. As shown in the Figure 13, t1~tn are the latency of each message, we use the average latency as the result.

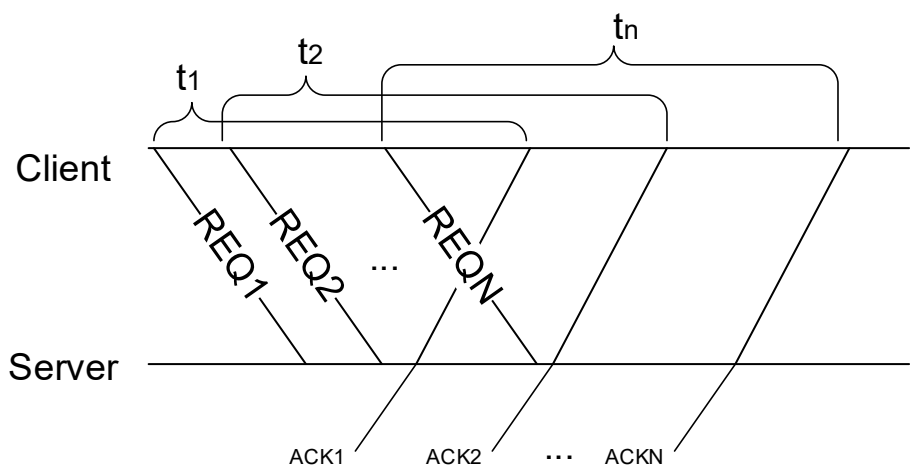

Figure 13 The sequence of test messages

Moreover, in order to compare the difference between LAN and LO, we have two different setups. For LAN group, the Server components and Client component are started in two different computers under the same LAN (100Mbps, Ethernet), and for LO group, the Server and Client components are started in the same physic server.

The result is shown in Figure 4. As shown in the figure, the communication delays under Link local are all under 18ms, and those under LAN are under 50ms. Since most of the messages used in actual projects sized from 64byte~2kbyte, the typical communication delay is around 10ms under LAN and 5ms under LO

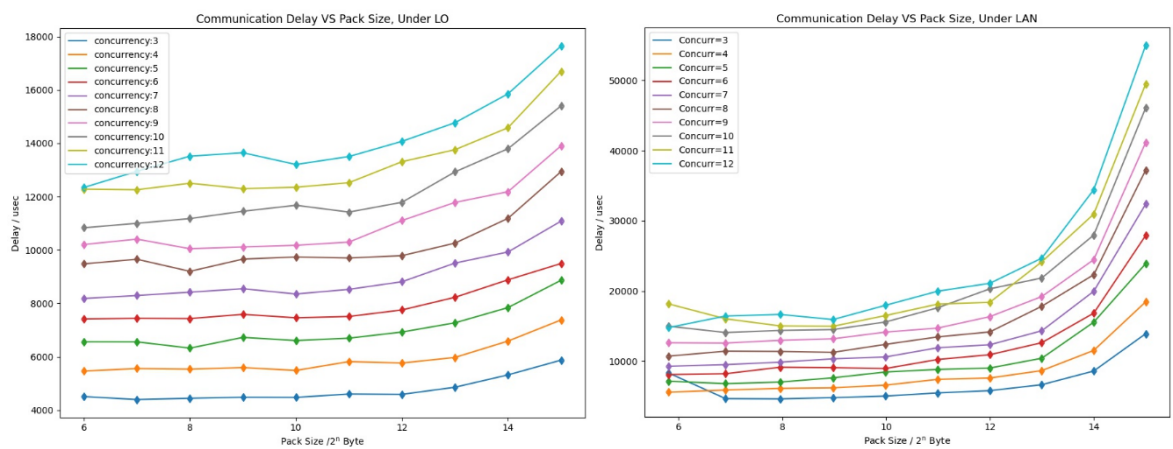

Figure 14 Communication delay VS Pack Size of RACS2, figure left shows the result under link local environment, figure right shows the result under 100M Local Area Network

As a comparison, we benchmarked the RTS2 framework in a similar way. The RTS2 has a limitation of 8k bytes for the size of message, so the result is not one-to-one correspondent to RACS2's.

Since the RTS2 framework uses the raw socket API, it out-performs the RACS2 for short message (size < 1k), but RACS2 shows advantage on long message. We also found that the latency of RTS2 spread more widely than RACS2, which means that RACS2 have better determinacy than RTS2.

## 5. APPLICATIONS

At present, RACS2 has been applied to several telescopes and observatories, including telescopes for Space Object observation, observational devices in Antarctic and control system of Wide Filed Survey Telescope Camera (WFCam).

### 8.1 SPACE OBJECT OBSERVATOIN

With more and more rockets, satellites and other spacecrafts enter the orbital environment, space debris is also increasing rapidly. Therefore, a number of small and medium-sized optical telescopes are planed around the world, to build a space debris monitoring and warning platform.

One of these telescopes is currently under construction. It has a reflecting telescope with diameter of 60cm and focal ratio of F/2.2. The system has a GPS module to provide time service, a serial port-based electronic focuser in the optical tube, an equatorial mount, an electronic dome controlled by serial port, a webcam, a weather station with web interface, and a panoramic camera.

It is also equipped with a PX4040 camera with a CMOS sensor size of 9um and resolution of 4Kx4K@12bit sampling, which is developed by our group. The frame rate can reach up to 10fps through USB3 interface. It is shown in Figure 15 .

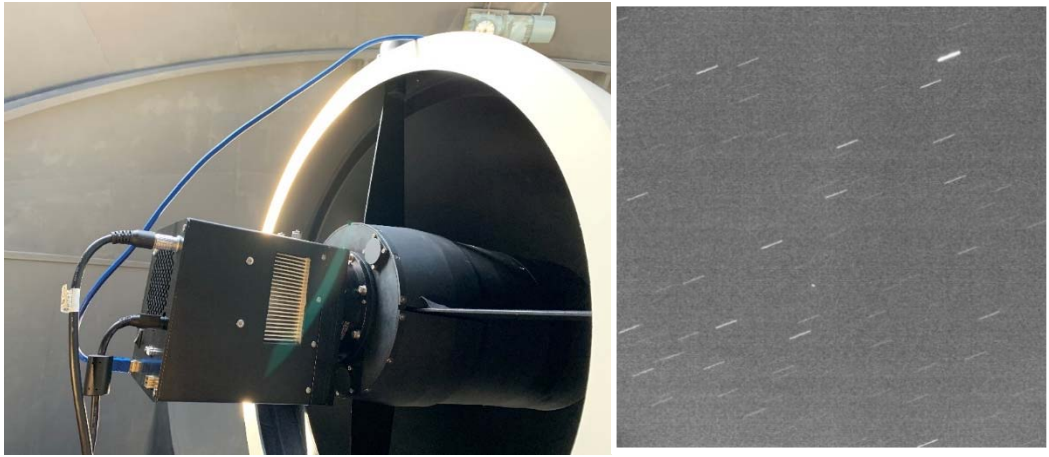

Figure 15 The telescope with PX4040 sCMOS camera and its image

We used the RACS2 to framework to develop the control and monitor software of this observatory. Every equipment, including the telescope mount, the dome, the camera, the weather station, the webcam and other devices, are connected to the RACS2 framework. Each subsystem can be access via the RACS2 User Interface.

The observation task of space debris is different from those of stellar. For the stellar observation, we send a target position to the mount, enable the tracking procedure, and start taking picture. But for space debris, we need to track the space debris by sending a set of positions to the mount and adjust the movement speed of each axis of the mount. RACS2 provides enough flexibility to accomplish such observations task. Because the observation tasks are just some hot pluggable scripts for RACS2, we also implement the stellar tracking, flat-field observation and other observation tasks. All these tasks can be called by users when needed.

This system has been running for more than one year and gathered a lot of data.

## 8.2 DOME A TWINS

Dome A Twins (DATs) telescope system is the second generation of Antarctic small telescope array after Chinese Small Telescope ARray (CSTAR). It includes two telescope each with a diameter of 145mm. Each telescope is equipped with a frame-transfer scientific CCD camera E4720 with a resolution of 1K * 1K. Since the whole telescope system will be installed at the dome A, Antarctica observatory, all equipment are specially designed for temperature as low as -80 ℃.

We build the control system of DATs based on RACS2. DATs needs to conduct long-term unattended observation in the harsh environment of Antarctica. The DATs has an equatorial mount, two CCD cameras and a power control subsystem. We already developed the basic control software of these equipment based on EPICS, so these EPICS IOCs are connected to RACS2 with EPICS-bridge.

The RACS2 system for DATs shares some features of space debris telescope. All observation tasks of DATs are stored in the observation database before deploying to the Antarctic, and then the observation tasks will be performed in the Antarctic for several months. Since it is very difficult to access the web interface due to the harsh network environment in the Antarctic, the operation towards

RACS2 can be done with command-line client via SSH session. Anyway, the system management and fault handling mechanism of RACS2 can ensure the system to survive from most simple exceptions. At present, DATs telescope with RACS2 system has been tested for several months at the Xuyi Station of the Purple Mountain Observatory As shown in Figure 16, it will be deployed to Dome A for scientific observation in the near future

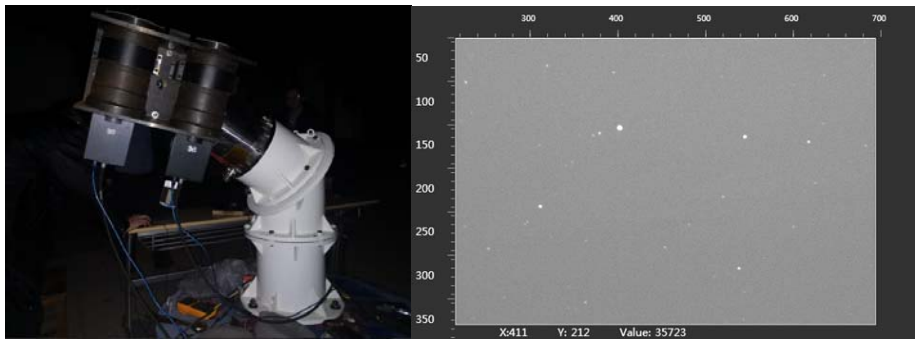

Figure 16 Field Test run of DATs

## 6. CONCLUSION

RACS2 is a modern distributed remote autonomous telescope control framework. Thanks to the decentralized distributed design, lightweight system structure, complete application scheme and excellent performance, RACS2 is lightweight and easy to use. Small teams can implement their product with RACS2 quiet easily. At present, RACS2 has been successfully applied to telescopes for Space Object observation, observational devices in Antarctic telescope and control system of Wide Filed Survey Telescope Camera (WFCam). Still, a lot of auxiliary function can be added to RACS2, and many general instruments shall be supported. Furthermore, we are developing many even more enthusiastic functions for RACS2, including the mechanism for the integration of multiple observatories, automatic observation based on AI, and special components for some specific astronomical applications, like time-domain astronomy and radio astronomy.